\documentclass[
    prd, twocolumn, superscriptaddress, nofootinbib, amsmath, amssymb, preprintnumbers,
]{revtex4-2}
\usepackage{booktabs}
\usepackage{threeparttable}
\usepackage{graphicx}
\usepackage{dcolumn}
\usepackage{bm}
\usepackage{mathtools}
\usepackage{ulem}
\usepackage{color,txfonts}
\usepackage{xcolor}
\definecolor{blue0}{rgb}{0,0,0.6}
\usepackage[colorlinks,linkcolor=blue0,anchorcolor=blue0,citecolor=blue0,urlcolor=blue0]{hyperref}

\begin{document}

\title{Multi-Period Fermi-LAT and LHAASO Constraints on Axion-Like Particles from Mrk 421 via $\chi^2$ Profile Likelihood with Gaussian Copula Correlation}

\author{Longhua Qin}
\affiliation{Department of Physics, Yuxi Normal University, Yuxi,Yunnan, 653100, People's Republic of China}
\email{lhz@yxnu.edu.cn}
\email{wangao@yxnu.edu.cn}

\author{Jiancheng Wang}
\affiliation{Yunnan Observatory, Chinese Academy of Sciences, Kunming, Yunnan, 650011, People's Republic of China}

\author{Chuyuan Yang}
\affiliation{Yunnan Observatory, Chinese Academy of Sciences, Kunming, Yunnan, 650011,  People's Republic of China}

\author{Huaizhen Li}
\affiliation{Department of Physics, Yuxi Normal University, Yuxi,Yunnan, 653100, People's Republic of China}

\author{Ao Wang}
\affiliation{Department of Physics, Yuxi Normal University, Yuxi,Yunnan, 653100, People's Republic of China}

\author{Weiwei Na}
\affiliation{Department of Physics, Yuxi Normal University, Yuxi,Yunnan, 653100, People's Republic of China}

\author{Hushan Xu}
\affiliation{Department of Physics, Yuxi Normal University, Yuxi,Yunnan, 653100, People's Republic of China}

\author{Zhongtao Han}
\affiliation{Department of Physics, Yuxi Normal University, Yuxi,Yunnan, 653100, People's Republic of China}

\author{Xiaogu Zhong}
\affiliation{College of Physics and Electronic Engineering, Qujing Normal University, Qujing 655011, People's Republic of China}

\author{Zunli Yuan}
\affiliation{Department of Physics and Synergistic Innovation Center for Quantum Effects and Applications, Hunan Normal University, Changsha,Hunan, 40081, People's Republic of China}

\author{Yubin Li}
\affiliation{School of Engineering and Technology, Baoshan University, Baoshan, 678000, People's Republic of China}

\author{Guangbo Long}
\affiliation{School of Intelligent Engineering, Shaoguan University, Shaoguan, Guangdong, 512005, People's Republic of China}
\email{727406989@qq.com}

\date{\today}

\begin{abstract}

We present a joint multi-period analysis of axion-like particles (ALPs) using five sets of simultaneous Fermi-LAT and LHAASO observations of the TeV blazar Mrk~421. Photon-ALP oscillations are calculated self-consistently together with EBL absorption for two representative jet emission models: a two-zone hybrid model and a single-zone hadronic model. To account for weak correlations among different observational epochs, we adopt a $\chi^2$-based profile analysis combined with a Gaussian copula ($\rho=0.03$). Nuisance parameters are optimized under the no-ALP hypothesis before profiling the ALP parameters $m_a$ and $g_{a\gamma}$. In the low-mass regime ($m_a\lesssim1$\,neV) we obtain a 95\% CL upper limit of $g_{a\gamma}=6.35\times10^{-13}\,\mathrm{GeV}^{-1}$. Over the full mass range $0.1$--$500$\,neV, the most conservative 95\% CL upper limits are $g_{a\gamma}<6.69\times10^{-12}\,\mathrm{GeV}^{-1}$ (two-zone) and $g_{a\gamma}<7.37\times10^{-12}\,\mathrm{GeV}^{-1}$ (single-zone). Variations of the EBL template, the Galactic and intergalactic magnetic-field models, the spectral binning, and residual instrumental systematics change these limits by only a few percent. The jet magnetic field is the leading systematic and can shift the low-mass bound by a factor of a few; only a substantially weaker single-zone field degrades the constraint by about an order of magnitude. The analysis framework developed here provides a statistically consistent approach for future ALP searches with multi-messenger gamma-ray data.

\end{abstract}

\pacs{Valid PACS appear here}

\maketitle

\section{\label{sec:Introduction}Introduction}

Axion-like particles (ALPs) are well-motivated extensions of the Standard Model that naturally arise in a broad class of theories beyond the Standard Model, including string-theory compactifications and other frameworks involving spontaneously broken global symmetries (e.g., see \citep{pec1977, svr2006, arv2010});
for cosmological aspects and early Universe production mechanisms (including the vacuum misalignment mechanism), see also \citep{pre1983,abb1983,din1983,vys1978,ber1991,ber1992,sak1994,sak1996,khl1999,khl2005,mar2016}. As light pseudo-scalar bosons coupled to photons through a two-photon interaction vertex, ALPs can induce photon-ALP oscillations in the presence of external magnetic fields, thereby producing potentially observable signatures in a variety of astrophysical environments \citep{sik1983, raf1988}. Owing to their weak couplings and light masses, ALPs are also considered viable dark matter candidates and may provide insight into several outstanding problems in particle physics and cosmology. Detailed studies of primordial axion field inhomogeneities, nonlinear modulations of the axion dark matter density distribution (including minicluster formation), and their cosmological implications can be found in \citep{sak1994,sak1996,khl1999,khl2005,kol1993,kol1994}, as well as in comprehensive reviews such as \citep{mar2016}.

Over the past decade, increasingly stringent limits on the ALP parameter space have been obtained from laboratory experiments, astrophysical observations, and cosmological probes. Helioscope experiments such as CAST have placed strong constraints on the ALP-photon coupling in the low-mass regime, excluding couplings above $g_{a\gamma}\sim 10^{-10},\mathrm{GeV}^{-1}$ for $m_a \lesssim 0.02$ eV \citep{cas2017}. More recently, gamma-ray observations of active galactic nuclei (AGNs), supernovae, galaxy clusters, and pulsars have substantially extended the accessible parameter space toward lower masses and weaker couplings \citep{mey2013B,aje2016,abe2024}. In particular, very high energy (VHE) gamma-ray observations have emerged as one of the most sensitive astrophysical probes of ultra-light ALPs in the neV mass range, where photon-ALP oscillations can occur efficiently over cosmological distances. Despite significant progress, large regions of parameter space relevant to both particle physics and cosmology remain unconstrained, motivating continued searches with next-generation gamma-ray instruments.

High energy gamma ray observations provide a particularly powerful probe of ALPs because gamma rays emitted by distant sources, such as blazars, may undergo photon-ALP oscillations during propagation through cosmic magnetic fields, including the Intergalactic Magnetic Field (IGMF). Such oscillations modify the photon survival probability and can generate characteristic spectral irregularities or spectral hardening features at high energies \citep{dea2007, mir2007, mey2013}. Consequently, observations of TeV blazars have been widely employed to search for ALP-induced effects and to derive constraints on the ALP parameter space.

A crucial ingredient in the propagation of VHE gamma rays is their interaction with the Extragalactic Background Light (EBL), which attenuates gamma rays through electron-positron pair production. The resulting optical depth introduces an energy-dependent exponential suppression of the observed photon flux at TeV energies. However, uncertainties in the EBL intensity and spectral shape remain at the level of \(\sim20\%\) in the wavelength range relevant for TeV \(\gamma\)-ray attenuation, and these uncertainties are known to be partially degenerate with photon-ALP oscillation effects \citep{dom2011,fra2017,gre2024}. In particular, a lower EBL normalization can mimic the spectral hardening expected from photon-ALP conversion, potentially leading to spurious indications of new physics if these effects are not modeled self-consistently.

While some studies adopted a factorized treatment in which photon-ALP oscillations and EBL absorption were computed independently and then combined multiplicatively (e.g., see \cite{mey2013,aje2016}), this approximation, although computationally efficient, neglects the nontrivial interplay between photon absorption and photon-ALP mixing during propagation and may therefore bias the inferred ALP constraints, particularly when EBL uncertainties become significant or when oscillation effects occur in strongly absorptive regimes. More recent analyses, including those based on the {\tt gammaALPs}\footnote{\url{https://github.com/me-manu/gammaALPs}} package \citep{mey2022} (e.g., see \citep{li2025,don2025}) and the seminal works by Galanti {\it et al.}~\citep{gal2020,gal2023a,gal2023b,gal2025}, employ the full transfer-matrix formalism that treats both effects simultaneously. In the present work we follow this approach.

On the observational side, recent advances in gamma-ray instrumentation have dramatically improved the sensitivity of ALP searches across a broad energy range. The Fermi-LAT provides precise spectral measurements in the GeV regime, while ground-based facilities such as LHAASO extend observations into the multi-TeV and PeV domains. The combination of these instruments enables broadband spectral studies with unprecedented sensitivity to energy-dependent distortions induced by photon-ALP oscillations. The blazar Mrk 421 is one of the brightest and most variable VHE emitters in the sky, offering a unique laboratory for probing both jet physics and fundamental physics beyond the Standard Model. Recent unbiased monitoring by LHAASO from March 2021 to March 2024 has revealed a clear dichotomy in its multiwavelength behavior: while the X-ray and VHE bands show a strong, lag-free correlation, the correlation between the high-energy (HE; GeV) band observed by Fermi-LAT and the VHE band is significantly weaker \citep{lha2026}. This is most evident in the long-term HE-to-VHE joint spectral energy distribution (SED) \citep{lha2026}, where the markedly larger variability amplitude in the VHE band compared with the GeV band, together with a harder-when-brighter VHE evolution, suggests that the two energy regimes originate from distinct physical processes or spatially separated regions within the jet \citep{lha2026}. Such a weak HE-VHE correlation provides strong motivation for adopting either a hadronic emission scenario or an energy-stratified two-component model when interpreting the broadband data for ALP searches.

In this work, we present a comprehensive search for ALPs using multi-epoch observations of the bright TeV blazar Mrk 421, combining broadband gamma-ray data from Fermi-LAT and LHAASO. We calculate the photon survival probability within a single-zone hadronic model and a two-zone hybrid model. Following \citep{qin2026}, photon-ALP mixing and EBL absorption are treated simultaneously in the propagation equation. Unlike the commonly adopted factorized approximation, the EBL attenuation is incorporated directly into the photon-ALP mixing matrix through an absorptive term in the propagation Hamiltonian. Adopting the EBL model of Saldana-Lopez et al.\ \citep{sal2021}, we further introduce an EBL normalization factor $k_{\rm EBL}$ to account for uncertainties in the EBL intensity during the likelihood analysis. To efficiently explore the multidimensional parameter space, we precompute the photon survival probability on a four-dimensional grid spanning photon energy, ALP parameters, and EBL normalization, and subsequently interpolate this grid during likelihood evaluation.

On the statistical side, most existing multi-epoch ALP searches treat different observational periods as completely independent and combine them through a direct summation of the per-epoch $\chi^2$ values (e.g., \cite{gao2024,li2024}). However, such an assumption neglects weak but potentially non-negligible residual correlations arising from shared source variability, common instrumental systematics, EBL-model uncertainties, and assumptions regarding cosmic magnetic-field configurations. Ignoring these effects may artificially tighten confidence intervals or introduce subtle biases into the inferred ALP constraints, particularly when the number of epochs is moderate and the datasets are partially correlated. To overcome this limitation, we introduce a Gaussian-copula formalism that explicitly models weak inter-epoch correlations through the transformed per-epoch $\Delta\chi^2$ excess distributions, adopting a conservative correlation coefficient of $\rho=0.03$. This approach preserves the $\chi^2$ profile likelihood framework while providing a statistically consistent treatment of correlated multi-period datasets. To derive robust and conservative constraints, we perform a joint $\chi^2$ profile likelihood analysis over multiple observational periods, allowing the intrinsic spectral parameters to vary independently between epochs while treating the ALP parameters as shared global quantities.

This paper is organized as follows. In Sec. II, we describe the photon-ALP propagation formalism, the modeling the ALP effect, and the statistical methodology, including the Gaussian copula treatment of inter-epoch correlations. In Sec. III, we present the results of the joint likelihood analysis and derive constraints on the ALP parameter space, comparing them with existing bounds from laboratory experiments such as CAST. Finally, we summarize our conclusions in Sec. IV.

\section{\label{sec:0}Methodology}
\subsection{Photon-ALP Propagation with EBL Absorption}
The propagation of VHE gamma rays in the presence of ALPs is governed by photon-ALP mixing in external magnetic fields (e.g., see \citep{raf1988,mir2007}). In this work, we model the propagation of photons emitted from the blazar Mrk 421 by solving the photon-ALP evolution equation along the entire propagation path, including the magnetic-field environments of the source jet and host system, the IGMF, and the Galactic magnetic field (GMF) of the Milky Way before the photons reach the observer.

The interaction between the ALP and photons can be described by the Lagrangian 
\begin{equation}
    \mathcal{L}_{a\gamma}=-\frac{1}{4}g_{a\gamma}F_{\mu\nu}\widetilde{F}^{\mu\nu}a=g_{a\gamma}a\vec{E}\cdot\vec{B},
\end{equation}
where $g_{a\gamma}$ is the coupling coefficient, $a$ represents the ALP field, $F_{\mu\nu}$ and $\widetilde{F}_{\mu\nu}$ are the electromagnetic tensor and its dual respectively, $\vec{E}$  is the photon electric field, and $\vec{B}$ is the magnetic field. 

The evolution of an photon-ALPs beam propagating through an external magnetic field along the $z$-axis is governed by a Schrödinger-like equation that accounts for mixing between photons and ALPs. The system is described by the state vector $\psi(z)\equiv(A_{x}, A_{y}, a)^T$

\begin{equation}
i \frac{d\psi(z)}{dz} = \mathcal{M} \psi(z),
\end{equation}

where $A_{x}$ and $A_{y}$ hold as two polarization amplitudes of the photon. 

The matrix $\mathcal{M}$ describes the ALP oscillations and the absorption of high-energy photons.

Assuming the transverse magnetic field $B_T$ parallel to $A_y$, $\mathcal{M}$ can be written as (eg., see\cite{lon2020,lon2021}) 
\begin{equation}
\label{eq:MM}
\mathcal{M}^{(0)} =   \begin{pmatrix}
    \Delta_{\perp}&0&0\\
    0&\Delta_{\parallel}&\Delta_{a\gamma}\\
    0&\Delta_{a\gamma}&\Delta_a\\
      \end{pmatrix},
\end{equation}  where the elements $\Delta_{\perp}$ and $\Delta_{\parallel}$ in matrix arise from photons propagating in a plasma and the QED vacuum polarisation effect (e.g., see \citep{qin2026}), respectively. They can be expressed as:
\begin{equation}
 \label{eq4_1}
    \Delta_{\parallel} = \Delta_{\rm pl} + \frac{7}{2}\Delta_{\rm QED} + \Delta_{\rm CMB}+ \frac{i}{2\lambda_\gamma},
  \end{equation}
\begin{equation}
 \label{eq4_2}
    \Delta_{\perp} = \Delta_{\rm pl} + 2\Delta_{\rm QED} + \Delta_{\rm CMB}+ \frac{i}{2\lambda_\gamma}.
  \end{equation}
  
For an electron density $n_{\rm e}$, the plasma contribution is given by $\Delta_{\rm pl} = -\omega_{\rm pl}/2E$, where the plasma frequency is defined as $\omega_{\rm pl}^2=4\pi e^2n_{\rm e}/m_e$. The QED vacuum-polarization contribution takes the form of $\Delta_{\rm QED} = \alpha E (B/B_{\rm cr})^2/45\pi$, where $\alpha$ is the fine-structure constant and $B_{\rm cr}\equiv m_e^2/e\simeq 4.4\times 10^{13}$ G denotes the critical magnetic field. In addition, the term $\Delta_{\rm CMB}=44\alpha^2E\rho_{\rm CMB}/(135m_e^4)$ describes the dispersive effect induced by photon-photon scattering on the cosmic microwave background (CMB) \citep{dob2015}, with the CMB energy density given by $\rho_{\rm CMB}=(\pi^2T^4)/15\sim 0.261 \rm {eVcm^{-3}}$. The rest of two terms, $\Delta_a$ and $\Delta_{a\gamma}$, characterize the ALP mass contribution and the photon-ALP mixing term, respectively, and are given by:

\begin{equation}
       \Delta_a = -\frac{m^2_a}{2E},
\end{equation}

\begin{equation}
       \Delta_{a\gamma} = \frac{g_{a\gamma}B}{2},
\end{equation}
where $m_a$ denotes the ALP mass. Gamma-ray photons emitted from TeV sources can continuously oscillate into ALPs and reconvert back into photons while propagating through different astrophysical environments, including the jet and host galaxy, the intergalactic medium, and the Milky Way. During propagation, VHE gamma rays are attenuated through interactions with the EBL. In the photon-ALP formalism, this absorption effect is incorporated directly into the mixing matrix $\mathcal{M}$ (see Eq.~\ref{eq:MM}) through the imaginary term $i/(2\lambda_\gamma)$, where $\lambda_\gamma$ represents the photon mean free path.

In this work, the EBL attenuation is modeled using the EBL template of \citep{sal2021}. Unlike the commonly adopted factorized approximation, the absorption effect is treated self-consistently within the photon-ALP propagation Hamiltonian rather than as an external multiplicative factor. To account for uncertainties in the overall EBL intensity, we introduce an EBL normalization factor $k_{\rm EBL}$, such that

\begin{equation}
\lambda_\gamma^{-1}\propto k_{\rm EBL}\cdot\tau,
\end{equation}

where $\tau$ is the optical depth predicted by the reference EBL model. The normalization factor is treated as a nuisance parameter in the likelihood analysis and is assigned a Gaussian prior,

\begin{equation}
\label{eq:GE}
k_{\rm EBL}\sim \mathcal{N}(\mu,\sigma^2),
\end{equation}

where $\mu=1.0$ and $\sigma=0.2$, corresponding to a ($\sim$20\%) uncertainty in the EBL normalization \citep{gre2024}.

\subsection{\label{Ma}Modeling the ALP Effect}

In this study we consider two representative jet scenarios motivated by existing multiwavelength models of Mrk 421: a two-zone hybrid configuration and a single-zone hadronic configuration. The two-zone case is intended to capture the coexistence of a compact, flaring hard-VHE component and a more extended, SSC-dominated baseline component (e.g., see \cite{sah2021}), rather than a unique, observationally reconstructed magnetic profile. In each component the jet field is treated as a superposition of helical and tangled contributions (e.g., see \citep{dig2023,kim2024}). The numerical values adopted below are not fitted to the present data set; they are representative parameters taken from published Mrk 421 jet models and are kept fixed in the photon--ALP propagation. Photon--ALP mixing is evaluated in this structured jet field from the adopted emission scale outward; the quoted \(r_{\mathrm{VHE}}\) therefore locates the emitting plasma and is not the full length of the mixing path. For the inner component we take \(B_0=0.03~\mathrm{G}\) and \(r_{\mathrm{VHE}}=0.005~\mathrm{pc}\) (e.g., see \citep{fra2023}), a compact scale of \(\sim1.5\times10^{17}~\mathrm{cm}\) consistent with the observed intra-day to few-day variability \citep{lha2026}. For the outer component we take a weaker field \(B_0=0.003~\mathrm{G}\) and \(r_{\mathrm{VHE}}=0.008~\mathrm{pc}\), representing a more extended SSC-like region. 

The second scenario is a single-zone hadronic configuration. Given the weak correlation between the GeV and TeV emission of Mrk 421, a hadronic origin of the high-energy component is physically plausible. We therefore compute photon--ALP mixing in a single jet environment whose magnetic field is again a superposition of helical and tangled components. We adopt a representative hadronic field strength \(B_0=1.0~\mathrm{G}\) and place the emitting plasma at \(r_{\mathrm{VHE}}=0.01~\mathrm{pc}\) from the black hole (e.g., see \cite{der2012,bot2013}); as in the two-component case, \(r_{\mathrm{VHE}}\) specifies the emission distance rather than the full mixing path, and the field parameters are kept fixed. 

We fully account for ALP-induced effects within the Galactic magnetic field and develop a self-consistent photon-propagation framework in which photon-ALP mixing and EBL absorption are treated simultaneously within the propagation equation \citep{qin2026}. For the blazar emission of Mrk 421, we adopt a typical Doppler boosting factor of 25 in our calculations (e.g., see \citep{qin2018}). For the intergalactic medium we adopt a stochastic-field model with electron-density normalization $n_0=10^{-7}\,\mathrm{cm}^{-3}$ and coherence length $L_0=1000\,\mathrm{pc}$. The IGMF strength is poorly constrained; current observations only provide an upper bound of $B_0\simeq1\,\mathrm{nG}$ (e.g., see \citep{dur2013,psh2016,alv2021,ria2024}). Photon-ALP mixing in this component is controlled by the product $g_{a\gamma}B$, so the uncertain field strength is degenerate with the coupling. We therefore consider the two limiting cases: $B_0=0$, which switches off intergalactic mixing, and $B_0=1\,\mathrm{nG}$, corresponding to that observational upper limit and hence to the strongest allowed IGMF mixing at a given $g_{a\gamma}$. The baseline is taken at $B_0=0$. The $B_0=1\,\mathrm{nG}$ scan is used as a comparison to the baseline and as a test of its robustness.

Subsequently, photon propagation through the IGMF and the GMF is calculated using the Jansson12 GMF model \citep{jan2012}. Based on the configurations described above, we precompute the corresponding photon survival probabilities with the public code {\tt gammaALPs}, which provides a comprehensive numerical framework for photon-ALP propagation in astrophysical magnetic fields \citep{mey2022}. These precomputed probabilities are then used in the subsequent spectral fitting and likelihood analyses.

\subsection{\label{SM analysis} Statistical Methodology}

To derive robust constraints on the ALP parameter space $(m_a,g_{a\gamma})$ from multi-epoch Fermi-LAT and LHAASO observations of Mrk~421, we employ a joint $\chi^2$ profile likelihood framework that incorporates the photon-ALP propagation model and a Gaussian copula to account for weak inter-epoch correlations.

The analysis is performed on the published SED flux points \citep{lha2026} rather than event-level data. Although photon counts follow Poisson statistics, the high-statistics nature of the Mrk~421 observations-approximately $1.33\times10^{5}$ gamma-like events and an overall significance of $\sim215\sigma$ in LHAASO WCDA places \citep{lha2026} the majority of bins in the regime where the Gaussian $\chi^2$ approximation is well justified. The multi-epoch SEDs and their uncertainties are shown in Fig.\ref{fig:bestModel}. We therefore adopt a profile-$\chi^2$ constructed from these published flux points. Residual mild non-Gaussianity in a few low-energy or low-flux bins is expected to be sub-dominant compared with the systematic uncertainties already included in the analysis.
For each epoch $p$, the agreement between the observed spectrum and the model is quantified using an asymmetric $\chi^2$ statistic that properly accounts for asymmetric statistical uncertainties:

\begin{flalign}
\chi^2_p(m_a, g_{a\gamma}, \boldsymbol{\nu}_p) = \sum_{i=1}^{M_p} \left( \frac{F_{i,\rm obs} - F_{i,\rm model}(m_a, g_{a\gamma}, \boldsymbol{\nu}_p)}{\sigma_i} \right)^2 + {\rm Pen}(\boldsymbol{\nu}_p),
\end{flalign}

where $F_{i,\rm model}$ denotes the predicted flux, including the photon survival probability $P_{\gamma\gamma}(E; m_a, g_{a\gamma}, k_{\rm EBL})$, and $\sigma_i$ represents the upper or lower uncertainty according to the sign of the residual. The nuisance-parameter vector $\boldsymbol{\nu}_p$ contains the intrinsic spectral parameters of epoch $p$ together with its corresponding EBL normalization factor $k_{\rm EBL}$. A Gaussian penalty term, ${\rm Pen}(\boldsymbol{\nu}_p)$, is imposed on $k_{\rm EBL}$ as defined in Eq.~(\ref{eq:GE}). The same prior distribution is adopted for all epochs, whereas the intrinsic spectral parameters are optimized independently for each epoch to accommodate temporal spectral variability.

To derive constraints on the ALP parameter space ($m_a,g_{a\gamma}$), we perform a joint $\chi^2$ profile likelihood analysis combining all observational epochs. At each fixed point in the $m_a$-$g_{a\gamma}$ plane, the nuisance parameters associated with each epoch are simultaneously optimized, yielding the corresponding $\chi^2$ profiled likelihood and test statistic.

Most previous multi-epoch ALP studies assume that different observational periods are statistically independent and therefore combine them through a direct summation of the individual $\chi^2$ values. However, this approximation neglects weak residual correlations that may arise from common source properties, instrumental systematics, EBL-model assumptions, and shared propagation effects. Although such correlations are expected to be small, they can nevertheless influence the resulting confidence regions when multiple datasets are combined.

To account for these effects, we employ a Gaussian-copula framework \citep{skl1959,nel2006,son2009,sat2011,joe2014,yua2018}, which provides a flexible description of weak dependence while preserving the marginal likelihood structure of each epoch. Since no direct empirical estimate of the inter-epoch correlation coefficient is available, we adopt a conservative correlation strength of $\rho=0.03$, corresponding to weak positive correlations at the few-percent level. This choice is motivated by potential common contributions from instrumental systematics, EBL-model assumptions, and source-related effects, while remaining sufficiently small to avoid introducing artificial correlations.

For each epoch $p$, we define the profiled excess statistic

\begin{equation}
\Delta\chi^2_p(m_a,g_{a\gamma})
=\chi^2_p(m_a,g_{a\gamma})-\chi^2_{p,\min},
\end{equation}

where $\chi^2_{p,\min}$ is obtained by minimizing $\chi^2_p$ with respect to the spectral nuisance parameters and the ALP parameters \((m_a,g_{a\gamma})\). Following the copula construction, the excess statistic is transformed into an approximately standard normal variable,

\begin{equation}
\zeta_p = {\rm sign}(\Delta\chi^2_p) \sqrt{|\Delta\chi^2_p|}.
\end{equation}

The joint test statistic is then written as

\begin{equation}
\chi^2_{\rm joint}(m_a, g_{a\gamma}) = \sum_{p=1}^{M_p} \chi^2_{p,\rm prof}(m_a, g_{a\gamma}) + C_{\rm copula},
\end{equation}

where the copula correction term is given by

\begin{equation}
C_{\rm copula} = \frac12 \left[ \ln(\det R) + \boldsymbol{\zeta}^{\rm T} (R^{-1} - I) \boldsymbol{\zeta} \right],
\end{equation}

here \(R\) denotes the correlation matrix entering the Gaussian copula, whose diagonal elements are unity and whose off-diagonal elements are set to the constant correlation coefficient \(\rho=0.03\) adopted in this work; \(I\) is the corresponding identity matrix.

Finally, the exclusion statistic is defined as

\begin{equation}
\label{eq:delta_chi2_excl}
\Delta\chi^2_{\rm excl}(m_a,g_{a\gamma})
=\chi^2_{\rm joint}(m_a,g_{a\gamma})-\chi^2_{\rm min},
\end{equation}
where $\chi^2_{\rm min}$ is the minimum of the joint statistic over the ALP parameter plane. A joint fit with both $m_a$ and $g_{a\gamma}$ free is driven to the lower boundary of $g_{a\gamma}$, i.e. the weak-mixing limit. In this regime $\chi^2_{\rm min}$ coincides with the no-ALP value $\chi^2_{\rm joint,0}$, so that Eq.~(\ref{eq:delta_chi2_excl}) is numerically identical to $\chi^2_{\rm joint}-\chi^2_{\rm joint,0}$. Following the convention adopted in previous gamma-ray ALP analyses, we quote two-dimensional 95\% confidence-level exclusions at $\Delta\chi^2_{\rm excl}=5.99$ and one-dimensional upper limits at \(\Delta\chi^2_{\rm excl}=2.71\) \citep{wil1938,cow1998}. 

\section{\label{sec:Result}Application and Results}

\subsection{Multi-epoch joint constraints}
For the ALP-photon oscillation analysis in this work, we use the long-term HE-to-VHE joint SED of Mrk 421 constructed from simultaneous Fermi-LAT and LHAASO observations spanning March 2021 to March 2024 \citep{lha2026}. The dataset contains five observational epochs. The weak GeV-TeV correlation and distinct variability amplitudes seen in the joint SED suggest that photons at these two energy bands are produced in separate emission regions, which motivates our adoption of a two-zone model \cite{sah2021}. Meanwhile, a single-zone hadronic model can also naturally explain the poor correlation between GeV and TeV photons \citep{der2012,bot2013}. Accordingly, we employ jet emission configurations typical of hadronic scenarios. Both models are implemented via the JetHelicalTangled module, after which we perform IGMF and GMF propagation using the Jansson12 model \citep{jan2012}.

The photon survival probability, $P_{\gamma\gamma}(E; m_a, g_{a\gamma}, k_{\rm EBL})$, is obtained by solving the full photon-ALP propagation equation along the entire line of sight. The observed spectral energy distribution is therefore modeled as

\begin{equation}
E^2 \frac{dN}{dE} = P_{\gamma\gamma}(E; m_a, g_{a\gamma}, k_{\rm EBL}) \times \phi_{\rm int}(E),
\end{equation}

where $\phi_{\rm int}(E)$ denotes the intrinsic source spectrum. Since the intrinsic spectral shape of TeV blazars is not known a priori, we first determine the preferred intrinsic spectrum independently of ALP-induced effects for each observational epoch of Mrk~421. Five commonly adopted spectral distributions are considered, including the power law (PL), broken power law (BPL), exponential cutoff power law (ECPL), log-parabola (LP), and log-parabola with exponential cutoff (LPC).

For each candidate model, we perform an independent fit under the null hypothesis of no photon-ALP mixing ($g_{a\gamma}=0$). The observed spectra are modeled using the corresponding intrinsic spectral function combined with the photon survival probability including EBL attenuation. The EBL attenuation is calculated using the template of \citep{sal2021}, while the EBL normalization factor $k_{\rm EBL}$ is treated as a free nuisance parameter introduced in Sec.~II.

Because the public Fermi-LAT and LHAASO spectral points correspond to finite energy bins, the model flux is not evaluated at the bin center but averaged over the bin. For the \(i\)th bin,
\begin{equation}
\left\langle
E^2\frac{dN}{dE}
\right\rangle_i
=\frac{1}{\Delta E_i}
\int_{E_{i,{\rm low}}}^{E_{i,{\rm high}}}
P_{\gamma\gamma}(E;m_a,g_{a\gamma},k_{\rm EBL})\,
\phi_{\rm int}(E)\,
\mathrm{d}E,
\end{equation}
where \(\Delta E_i=E_{i,{\rm high}}-E_{i,{\rm low}}\). The same bin average is used both in the no-ALP model selection and in the subsequent ALP profile scan, so that the comparison with the measured SED points employs a consistent energy weighting.

\begin{figure*}
\centering
\includegraphics[width=\textwidth]{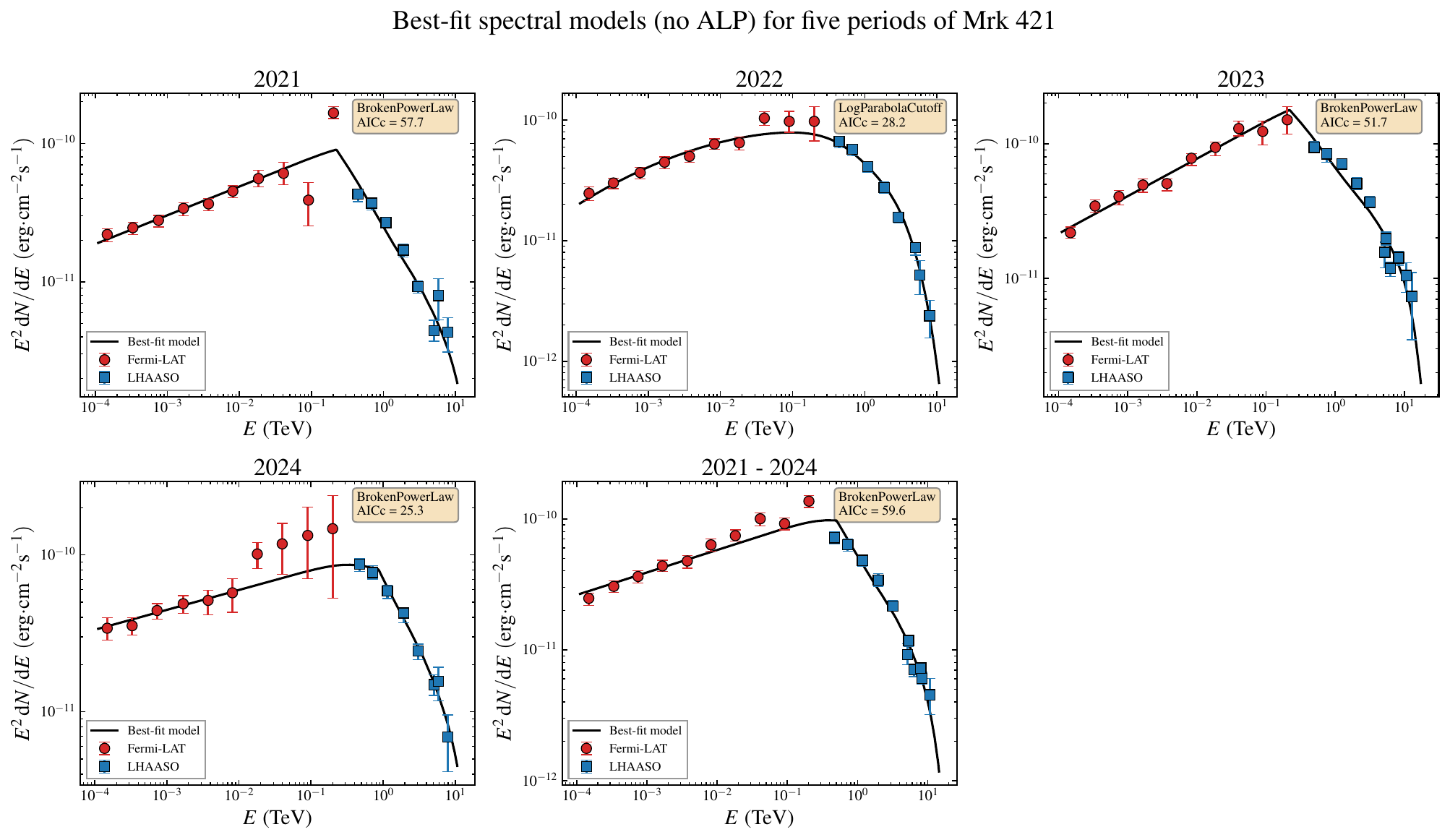}
\caption{Best-fit spectral energy distributions (no ALP) for the five observational periods of Mrk~421.
Red circles and blue squares denote the Fermi-LAT and LHAASO data points, respectively. Solid green curves show the best-fit spectral model for each period, selected by the AICc. The model name and corresponding AICc value are indicated in each panel.}
\label{fig:bestModel}
\end{figure*}

To avoid overfitting while preserving sufficient spectral flexibility, the preferred intrinsic spectral model for each epoch is selected via the Akaike Information Criterion corrected for small sample sizes (AICc) \citep{aka1974,gut2003}:

\begin{equation}
{\rm AICc} = \chi^2_{\rm min} + 2k + \frac{2k(k+1)}{n-k-1},
\end{equation}

where $k$ denotes the number of free model parameters and $n$ is the number of spectral data points. AICc balances goodness-of-fit against model complexity and applies a stronger penalty for additional parameters when the number of spectral data points is limited, which is particularly suitable for the modest sample sizes in each observational period of this work. The spectral model with the minimum AICc is adopted as the preferred baseline intrinsic spectrum for the subsequent ALP profile-likelihood analysis. The resulting best-fit spectral energy distributions for the five observational periods are shown in Fig.~\ref{fig:bestModel}.

We explored the ALP parameter space using a two-dimensional logarithmic grid. The ALP mass $m_a$ was sampled with 100 points spanning from 0.1 to 500 neV, and the photon-ALP coupling $g_{a\gamma}$ with 280 points spanning from $5\times10^{-14}$ to $10^{-7}\mathrm{GeV}^{-1}$. This sampling provides adequate resolution for determining the corresponding confidence intervals and exclusion contours.

The photon survival probability entering the likelihood analysis was precomputed on a four-dimensional grid in energy, photon-ALP coupling, ALP mass, and EBL normalization. Unlike searches dominated by stochastic spectral irregularities induced by turbulent IGMFs (e.g., see \cite{lon2020,lon2021}), the propagation distance of Mrk 421 is short, so photon-ALP conversion in the intergalactic medium is subdominant relative to the jet and Galactic contributions. We therefore use the ensemble-averaged transfer probability. The same pipeline is kept for all scans. At the baseline $B_{\rm IGMF}=0$ the random-field average is redundant, because every realization is identical; the $N_{\rm sim}=10$ average and the $N_{\rm sim}=300$ test matter only for the $1\,\mathrm{nG}$ comparison. The survival probability is then smoothed in logarithmic energy with a one-dimensional Gaussian filter corresponding to a 15\% relative energy resolution, thereby removing oscillatory features that cannot be resolved by Fermi-LAT or LHAASO.

\begin{figure*}
\centering
\includegraphics[width=\textwidth]{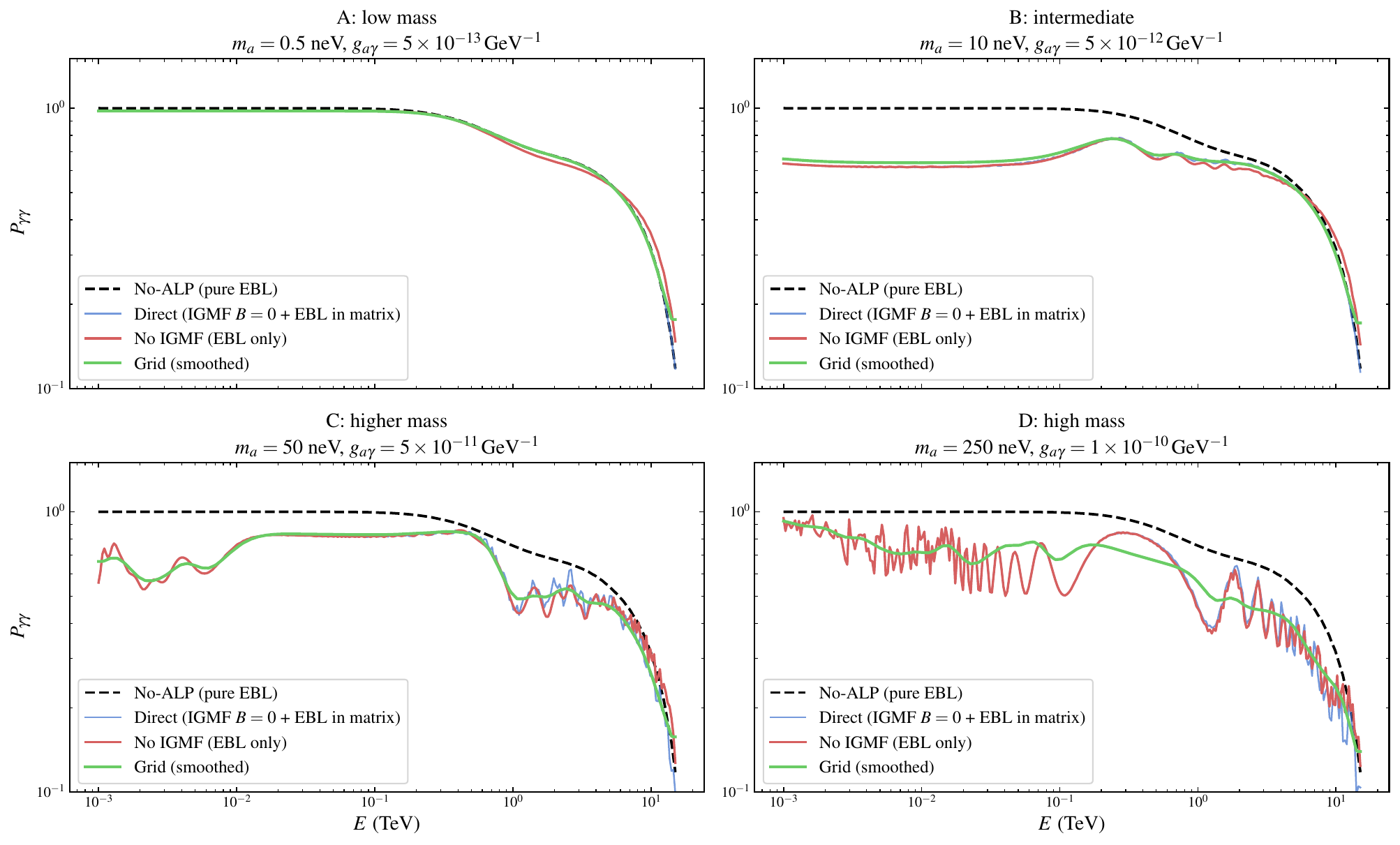}
\caption{Photon survival probability for Mrk 421 at selected $(m_a,g_{a\gamma})$ points with $k_{\rm EBL}=1.0$. The pure EBL attenuation is shown together with the direct {\tt gammaALPs} calculation including the IGMF module at $B_{\rm IGMF}=0$ and the result from the precomputed grid (ensemble average over $N_{\rm sim}=10$, followed by 15\% Gaussian energy-resolution smoothing). Also shown is the same jet+GMF setup with the IGMF module removed and replaced by the Saldana-Lopez EBL optical-depth module (no mixing, absorption only). A test with $N_{\rm sim}=300$ leaves the smoothed curves unchanged.}
\label{fig:survival_comparison}
\end{figure*}

Figure~\ref{fig:survival_comparison} compares, within the two-zone hybrid jet baseline, four calculations at representative points spanning low to high ALP mass: $(m_a,g_{a\gamma})=(0.5\,\mathrm{neV},\,5\times10^{-13}\,\mathrm{GeV}^{-1})$, $(10\,\mathrm{neV},\,5\times10^{-12}\,\mathrm{GeV}^{-1})$, $(50\,\mathrm{neV},\,5\times10^{-11}\,\mathrm{GeV}^{-1})$, and $(250\,\mathrm{neV},\,1\times10^{-10}\,\mathrm{GeV}^{-1})$. At this redshift the direct {\tt gammaALPs} curve with the IGMF module retained but $B_{\rm IGMF}=0$ is almost indistinguishable from the curve obtained by removing the IGMF module and replacing it with Saldana-Lopez EBL absorption only. Both remain distinct from the no-ALP EBL attenuation, as expected from mixing in the jet and in the Galaxy. This near coincidence is specific to a nearby source such as Mrk 421 and should not be extrapolated to high redshift, where a finite IGMF and in-matrix EBL absorption can accumulate over a much longer path. We therefore set $B_{\rm IGMF}=0$ in the baseline: numerically this is equivalent to an EBL-only intergalactic segment, while mixing and EBL absorption are still evolved together in the transfer matrix. The same setup can be reused without change for high-redshift sources, for which dropping the IGMF module would no longer be justified. The likelihood grid uses 15\% Gaussian energy-resolution smoothing and matches the direct calculation once unresolved oscillations have been averaged out.

\begin{figure}[h]
\includegraphics[scale=0.23]{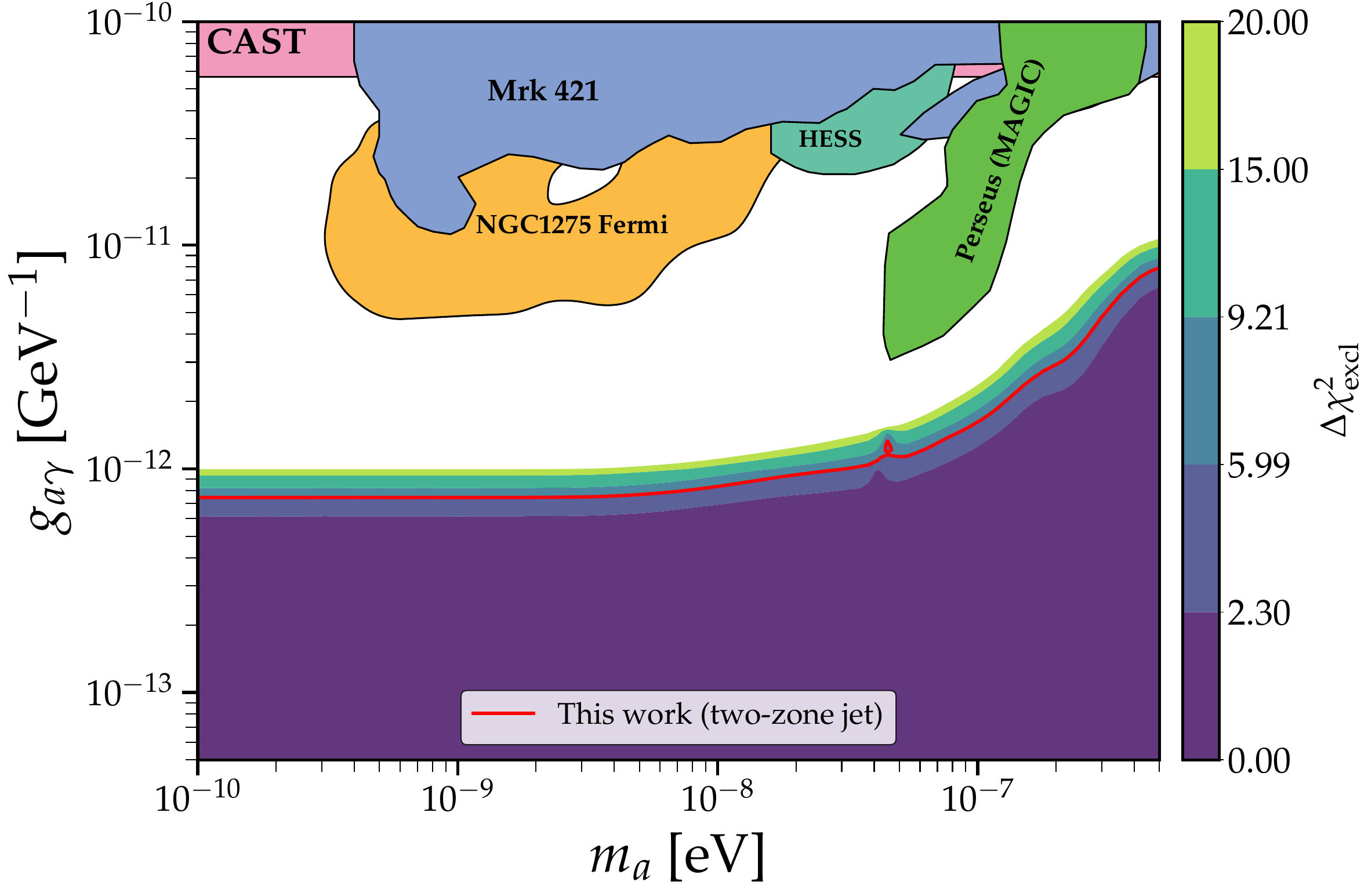}
\caption{Exclusion limits on the ALP-photon coupling $g_{a\gamma}$ versus the ALP mass $m_a$ obtained in this work using the two-zone hybrid model for Mrk 421 (red line, 95\% CL). The color scale shows the excluded $\Delta\chi^2$ values. For comparison, we also display the constraints established by the CAST experiment \citep{cas2017}, the H.E.S.S. observations of PKS 2155-304 \citep{abr2013}, the Fermi-LAT observation of NGC 1275 \citep{aje2016}, the ARGO-Fermi observation of Mrk 421\citep{li2021} and the MAGIC observation of Perseus cluster of galaxies \citep{ale2010}.}
    \label{fig:twozone}
\end{figure}

Fig.~\ref{fig:twozone} presents the 95\% confidence-level (CL) constraints on the ALP mass \(m_a\) and photon-ALP coupling \(g_{a\gamma}\) obtained from the profile-likelihood analysis in the two-zone hybrid scenario, using bin-averaged model fluxes. At low masses (\(m_a\lesssim1~\mathrm{neV}\)) the two-dimensional exclusion is nearly mass-independent, with \(g_{a\gamma}>7.34\times10^{-13}~\mathrm{GeV}^{-1}\) ruled out. The corresponding conservative one-dimensional limit in this window is \(g_{a\gamma}<6.35\times10^{-13}~\mathrm{GeV}^{-1}\). As \(m_a\) increases, the upper boundary of the exclusion rises through the discrete steps set by the coupling grid, reaching the \(10^{-12}~\mathrm{GeV}^{-1}\) level near \(m_a\sim30~\mathrm{neV}\). For \(m_a\gtrsim100~\mathrm{neV}\) the sensitivity degrades rapidly, and the two-dimensional upper edge attains \(7.78\times10^{-12}~\mathrm{GeV}^{-1}\) at \(m_a=500~\mathrm{neV}\). The exclusion therefore retains the standard photon-ALP pattern: a narrow, almost mass-insensitive window at low mass and a progressive relaxation of the upper bound toward larger masses. Across the full scanned range \(0.1\)–\(500~\mathrm{neV}\), the most conservative one-dimensional 95\% CL limit is \(g_{a\gamma}<6.69\times10^{-12}~\mathrm{GeV}^{-1}\).

\begin{figure}[h]
\includegraphics[scale=0.23]{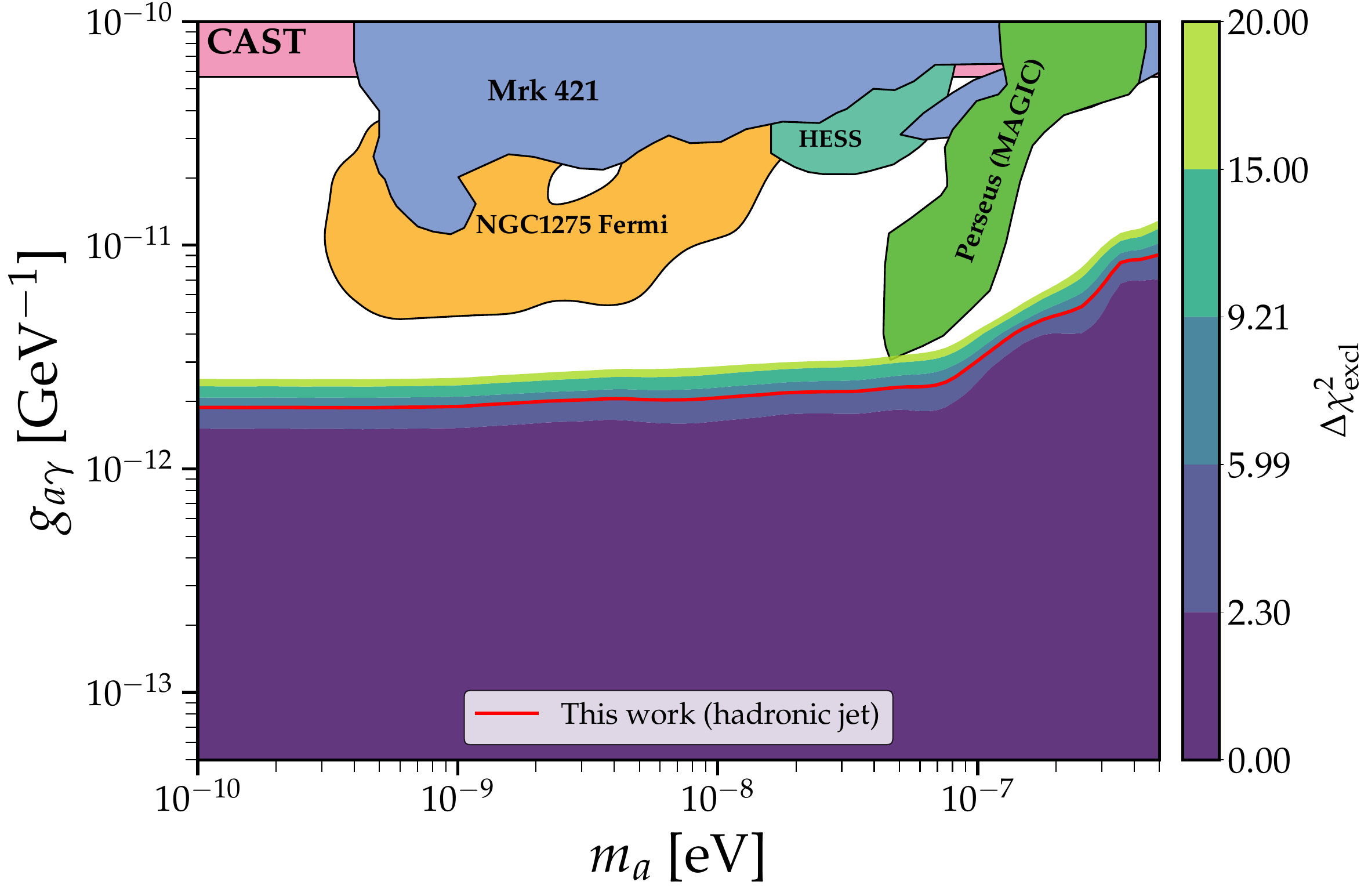}
\caption{Exclusion limits on the ALP-photon coupling $g_{a\gamma}$ versus the ALP mass $m_a$ obtained in this work using the single-zone hadronic model for Mrk 421.}
    \label{fig:onezone}
\end{figure}

The corresponding 95\% CL constraints for the single-zone hadronic model are shown in Fig.~\ref{fig:onezone}. The same bin-averaged profile-likelihood analysis is applied over \(0.1\leq m_a\leq500~\mathrm{neV}\) and yields the same qualitative mass dependence as in the two-zone case. At low masses (\(m_a\lesssim1~\mathrm{neV}\)) the two-dimensional exclusion remains nearly flat, with \(g_{a\gamma}>1.86\times10^{-12}~\mathrm{GeV}^{-1}\) ruled out. The associated one-dimensional limit in this window is \(g_{a\gamma}<1.57\times10^{-12}~\mathrm{GeV}^{-1}\). This weak mass dependence follows because, well below the resonant regime, the conversion probability is controlled mainly by \(g_{a\gamma}\) rather than by \(m_a\). Above \(1~\mathrm{neV}\) the upper boundary of the exclusion rises, with the discrete steps reflecting the coupling grid and the energy-dependent sensitivity of the SED. For \(m_a\gtrsim100~\mathrm{neV}\) the constraints weaken rapidly, and the two-dimensional upper edge reaches \(8.73\times10^{-12}~\mathrm{GeV}^{-1}\) at \(m_a=500~\mathrm{neV}\). Across the full mass range the most conservative one-dimensional 95\% CL limit is \(g_{a\gamma}<7.37\times10^{-12}~\mathrm{GeV}^{-1}\).

\subsection{Robustness against magnetic-field assumptions and EBL modelling}

The photon-ALP conversion probability depends on the magnetic-field configuration along the line of sight and on the normalisation of the extragalactic background light. We examine the principal systematic uncertainties associated with these ingredients below. 

To quantitatively assess the robustness of the derived ALP constraints under different astrophysical assumptions (e.g., alternative magnetic-field configurations or EBL models), we compare the corresponding 95\% CL upper-limit curves, $g_{95}(m_a)$, on a point-by-point basis over the full scanned ALP mass range. For each pair of configurations we define the ratio
\begin{equation}
R(m_a)=\frac{g_{95}^{(1)}(m_a)}{g_{95}^{(2)}(m_a)},
\end{equation}
where the superscripts denote the two scenarios being compared.

Since exclusion limits are conventionally presented in logarithmic parameter space, we further quantify the overall difference between two exclusion curves by the root-mean-square (RMS) logarithmic deviation,
\begin{equation}
\mathrm{RMS}_{\log}
=
\sqrt{
\frac{1}{N}
\sum_{i=1}^{N}
\left[
\log_{10}g_{95}^{(1)}(m_i)
-
\log_{10}g_{95}^{(2)}(m_i)
\right]^2
},
\end{equation}
where $N$ is the number of sampled ALP masses. The quantity $\mathrm{RMS}_{\log}$ is expressed in decades (dex) and provides a compact measure of the average relative deviation between two exclusion curves over the entire parameter space. Smaller values of $\mathrm{RMS}_{\log}$ indicate a higher degree of agreement and therefore stronger robustness of the derived constraints against the adopted astrophysical assumptions.

This logarithmic comparison is motivated by the standard presentation of exclusion limits in ALP and dark-matter searches, where limits typically span several orders of magnitude and are displayed on logarithmic scales (e.g,. see \citep{bax2021,gao2024}). Rather than relying on visual inspection of nearly overlapping exclusion curves, the RMS logarithmic deviation provides a concise quantitative metric for evaluating the stability of the constraints under different systematic assumptions.

\subsubsection{Uncertainty of the jet magnetic field}

Table~\ref{tab:jetB} summarises the 95\% CL upper limits obtained when the jet magnetic-field strength is varied for both emission models.

\begin{table*}[t]
\centering
\renewcommand{\arraystretch}{1.35}
\caption{Conservative 95\% CL upper limits on $g_{a\gamma}$ for different jet magnetic-field configurations.}
\label{tab:jetB}
\begin{tabular}{l@{\hspace{2.3em}}l@{\hspace{2.3em}}l@{\hspace{2.3em}}c}
\hline
Model & Jet magnetic field & Full mass range $g_{a\gamma}$ (95\% CL) &  $m_a <$ 1.0~neV~$g_{a\gamma}$ (95\% CL) \\
\hline
Two-zone (baseline) & $B_{\rm in}=0.03\,{\rm G}$, $B_{\rm out}=0.003\,{\rm G}$ & $<6.69\times10^{-12}\,{\rm GeV}^{-1}$ & $<6.35\times10^{-13}\,{\rm GeV}^{-1}$ \\
Two-zone (stronger) & $B_{\rm in}=0.1\,{\rm G}$, $B_{\rm out}=0.03\,{\rm G}$ & $<4.63\times10^{-12}\,{\rm GeV}^{-1}$ & $<9.88\times10^{-14}\,{\rm GeV}^{-1}$ \\
Two-zone (weaker) & $B_{\rm in}=0.01\,{\rm G}$, $B_{\rm out}=0.001\,{\rm G}$ & $<3.86\times10^{-12}\,{\rm GeV}^{-1}$ & $<1.63\times10^{-12}\,{\rm GeV}^{-1}$ \\
Single-zone (baseline) & $B=1\,{\rm G}$ & $<7.37\times10^{-12}\,{\rm GeV}^{-1}$ & $<1.57\times10^{-12}\,{\rm GeV}^{-1}$  \\
Single-zone (stronger) & $B=2\,{\rm G}$ & $<5.90\times10^{-12}\,{\rm GeV}^{-1}$ & $<1.60\times10^{-13}\,{\rm GeV}^{-1}$  \\
Single-zone (weaker) & $B=0.1\,{\rm G}$ & $<5.66\times10^{-11}\,{\rm GeV}^{-1}$ & $<1.60\times10^{-11}\,{\rm GeV}^{-1}$  \\
\hline
\end{tabular}
\end{table*}

The jet magnetic field is the leading systematic, because mixing in the emission region scales with $g_{a\gamma}B$. For the two-zone baseline ($B_{\rm in}=0.03\,\mathrm{G}$, $B_{\rm out}=0.003\,\mathrm{G}$), lowering both components to $B_{\rm in}=0.01\,\mathrm{G}$ and $B_{\rm out}=0.001\,\mathrm{G}$ loosens the low-mass bound from $6.35\times10^{-13}$ to $1.63\times10^{-12}\,\mathrm{GeV}^{-1}$ (mean ratio 2.12, RMS $0.35\,\mathrm{dex}$, almost entirely from $m_a\lesssim10\,\mathrm{neV}$). Raising the fields to $B_{\rm in}=0.1\,\mathrm{G}$ and $B_{\rm out}=0.03\,\mathrm{G}$ tightens the same window to $9.88\times10^{-14}\,\mathrm{GeV}^{-1}$. We do not adopt that tighter number as the published limit: the baseline fields are representative values for Mrk 421, not the configuration that maximises the ALP constraint. The weaker pair, especially $B_{\rm out}=0.001\,\mathrm{G}$, also lies below the range commonly inferred for the compact VHE zone of HBLs (e.g., see \citep{fra2023,bot2013}), so it is used only as a robustness test.

The full-range conservative figure moves the other way for the weaker field, from $6.69\times10^{-12}$ to $3.86\times10^{-12}\,\mathrm{GeV}^{-1}$. That quantity is simply the largest $g_{95}(m_a)$ on the grid and is set by the high-mass tail, where a smaller $\Delta_{a\gamma}$ changes the rapidly oscillating regime after energy-resolution smoothing. It is not a physically preferred bound. The relevant measure of sensitivity remains the low-mass window, which follows the expected $g_{a\gamma}B$ scaling.

For the single-zone hadronic model the baseline field is \(B=1\,\mathrm{G}\). Reducing it to \(0.1\,\mathrm{G}\) degrades the sensitivity by about an order of magnitude: the full-range limit loosens from \(7.37\times10^{-12}\) to \(5.66\times10^{-11}\,\mathrm{GeV}^{-1}\), and the \(m_a<1\,\mathrm{neV}\) limit from \(1.57\times10^{-12}\) to \(1.60\times10^{-11}\,\mathrm{GeV}^{-1}\). A point-by-point comparison gives a mean ratio of 9.44 and an RMS logarithmic difference of \(0.97\,\mathrm{dex}\). Increasing the field to \(2\,\mathrm{G}\) tightens the full-range limit to \(5.90\times10^{-12}\,\mathrm{GeV}^{-1}\) and the low-mass window to \(1.60\times10^{-13}\,\mathrm{GeV}^{-1}\). The mean ratio over the whole curve is 0.22 (RMS \(0.82\,\mathrm{dex}\)), so both the low-mass bound and the intermediate-mass part of the curve become stronger.

In both emission scenarios the jet field is the dominant systematic (Table~\ref{tab:jetB}). Within a factor of a few of the two-zone baseline the low-mass limit moves by a factor of a few, as expected from \(g_{a\gamma}B\) scaling. The single-zone case is more sensitive to \(B\): a substantially weaker field (\(0.1\,\mathrm{G}\)) degrades the limit by an order of magnitude, while raising the field from \(1\,\mathrm{G}\) to \(2\,\mathrm{G}\) tightens the \(m_a<1\,\mathrm{neV}\) bound by a similar factor. Because the adopted baseline values lie in the range inferred from multi-wavelength modelling of Mrk 421, we take the reported limits as representative of that modelling, while noting that a much weaker single-zone field would weaken them significantly.

\subsubsection{Uncertainty of the intergalactic magnetic field}

\begin{table*}[t]
\centering
\renewcommand{\arraystretch}{1.35}
\caption{Effect of the intergalactic magnetic-field strength (all other parameters fixed to the baseline configuration).}
\label{tab:igmf}
\begin{tabular}{l@{\hspace{4.9em}}l@{\hspace{4.9em}}l@{\hspace{4.5em}}c@{\hspace{4.5em}}l}
\hline
Model & $B_{\rm IGMF}$ & Full mass range $g_{a\gamma}$ (95\% CL) &  $m_a <$ 1.0~neV~$g_{a\gamma}$ (95\% CL) \\
\hline
Two-zone & $0\,{\rm nG}$ (baseline) & $<6.69\times10^{-12}\,{\rm GeV}^{-1}$ & $<6.35\times10^{-13}\,{\rm GeV}^{-1}$ \\
Two-zone & $1\,{\rm nG}$ & $<6.65\times10^{-12}\,{\rm GeV}^{-1}$ & $<6.27\times10^{-13}\,{\rm GeV}^{-1}$ \\
Single-zone & $0\,{\rm nG}$ (baseline) & $<7.37\times10^{-12}\,{\rm GeV}^{-1}$ & $<1.57\times10^{-12}\,{\rm GeV}^{-1}$  \\
Single-zone & $1\,{\rm nG}$ & $<7.37\times10^{-12}\,{\rm GeV}^{-1}$ & $<1.56\times10^{-12}\,{\rm GeV}^{-1}$  \\
\hline
\end{tabular}
\end{table*}

The intergalactic magnetic field is poorly constrained: present observations provide only an upper bound of order \(1\,\mathrm{nG}\). Our baseline analyses adopt \(B_{\rm IGMF}=0\). To quantify the associated uncertainty we repeated the full multi-period \(\chi^2\) profile-likelihood scan at that observational upper limit, \(B_{\rm IGMF}=1\,\mathrm{nG}\).

For the two-zone hybrid scenario the conservative full-range limit changes from \(6.69\times10^{-12}\,{\rm GeV}^{-1}\) to \(6.65\times10^{-12}\,{\rm GeV}^{-1}\), and the limit for \(m_a<1\,{\rm neV}\) from \(6.35\times10^{-13}\,{\rm GeV}^{-1}\) to \(6.27\times10^{-13}\,{\rm GeV}^{-1}\). A point-by-point comparison of the two one-dimensional exclusion curves yields a mean ratio of 0.987 and an RMS logarithmic difference of \(0.008\,{\rm dex}\). For the single-zone hadronic scenario the full-range limit remains \(7.37\times10^{-12}\,{\rm GeV}^{-1}\), while the low-mass limit shifts only from \(1.57\times10^{-12}\,{\rm GeV}^{-1}\) to \(1.56\times10^{-12}\,{\rm GeV}^{-1}\), with a mean ratio of 0.997 and an RMS difference of \(0.002\,{\rm dex}\). The one-dimensional upper limits are summarized in Table~\ref{tab:igmf}. This near-insensitivity is expected for a low-redshift source such as Mrk 421 (\(z=0.031\)), where the path length available for IGMF-induced oscillations is short. The precise value of the IGMF within the observationally allowed window therefore has a negligible impact on the final ALP constraints.

\subsubsection{Dependence on Galactic magnetic-field models}

\begin{table*}[t]
\centering
\renewcommand{\arraystretch}{1.35}
\caption{Conservative 95\% CL upper limits on $g_{a\gamma}$ obtained with different Galactic magnetic-field models (all other parameters fixed to the baseline configuration).}
\label{tab:gmf}
\begin{tabular}{l@{\hspace{3.0em}}l@{\hspace{3.0em}}l@{\hspace{3.0em}}c@{\hspace{3.0em}}l}
\hline
Model & GMF model & Full mass range $g_{a\gamma}$ (95\% CL) &  $m_a <$ 1.0~neV~$g_{a\gamma}$ (95\% CL) \\
\hline
Two-zone & Jansson \& Farrar (2012) (baseline) & $<6.69\times10^{-12}\,{\rm GeV}^{-1}$ & $<6.35\times10^{-13}\,{\rm GeV}^{-1}$ \\
Two-zone & Unger \& Farrar (2023), base & $<6.69\times10^{-12}\,{\rm GeV}^{-1}$ & $<6.36\times10^{-13}\,{\rm GeV}^{-1}$\\
Single-zone & Jansson \& Farrar (2012) (baseline) & $<7.37\times10^{-12}\,{\rm GeV}^{-1}$ & $<1.57\times10^{-12}\,{\rm GeV}^{-1}$  \\
Single-zone & Unger \& Farrar (2023), base & $<7.37\times10^{-12}\,{\rm GeV}^{-1}$ & $<1.57\times10^{-12}\,{\rm GeV}^{-1}$\\
\hline
\end{tabular}
\end{table*}

Photon–ALP mixing in the Milky Way is included in the propagation calculation. Because Mrk 421 lies at high Galactic latitude, the line-of-sight path through the Galactic magnetic field is relatively short, and the GMF contribution is expected to be sub-dominant compared with the jet field. Nevertheless, to quantify any residual model dependence we repeated the full multi-period analysis using the more recent Unger \& Farrar (2023) “base” model \citep{ung2024} in place of the baseline Jansson \& Farrar (2012) model \citep{jan2012}.

For the two-zone hybrid scenario the conservative full-range limit remains \(g_{a\gamma}<6.69\times10^{-12}\,{\rm GeV}^{-1}\), and the limit for \(m_a<1\,{\rm neV}\) changes only from \(6.35\times10^{-13}\,{\rm GeV}^{-1}\) to \(6.36\times10^{-13}\,{\rm GeV}^{-1}\). A point-by-point comparison of the two one-dimensional exclusion curves yields a mean ratio of 1.000 and an RMS logarithmic difference of \(0.0008\,{\rm dex}\). For the single-zone hadronic scenario both the full-range limit (\(7.37\times10^{-12}\,{\rm GeV}^{-1}\)) and the low-mass limit (\(1.57\times10^{-12}\,{\rm GeV}^{-1}\)) are unchanged, with a mean ratio of 1.000 and an RMS difference of \(0.0018\,{\rm dex}\). The one-dimensional upper limits in Table~\ref{tab:gmf} confirm that the choice between these two widely used coherent GMF models has a negligible impact on the ALP constraints for this high-latitude source. We therefore regard the baseline limits as robust against the present uncertainty in the large-scale Galactic magnetic field.

\subsubsection{Sensitivity to the EBL model}

\begin{table*}[t]
\centering
\renewcommand{\arraystretch}{1.35}
\caption{Conservative 95\% CL upper limits on $g_{a\gamma}$ obtained with different EBL models (all other parameters fixed to the baseline configuration).}
\label{tab:ebl}
\begin{tabular}{l@{\hspace{3.8em}}l@{\hspace{3.8em}}l@{\hspace{3.8em}}c@{\hspace{3.8em}}l}
\hline
Model & EBL model & Full mass range $g_{a\gamma}$ (95\% CL) &  $m_a <$ 1.0~neV~$g_{a\gamma}$ (95\% CL) \\
\hline
Two-zone & Saldana-Lopez et al.\ (baseline) & $<6.69\times10^{-12}\,{\rm GeV}^{-1}$ & $<6.35\times10^{-13}\,{\rm GeV}^{-1}$ \\
Two-zone & Franceschini et al.\ & $<6.80\times10^{-12}\,{\rm GeV}^{-1}$ &  $<6.52\times10^{-13}\,{\rm GeV}^{-1}$ \\
Single-zone & Saldana-Lopez et al.\ (baseline) & $<7.37\times10^{-12}\,{\rm GeV}^{-1}$ & $<1.57\times10^{-12}\,{\rm GeV}^{-1}$  \\
Single-zone & Franceschini et al.\ & $<7.55\times10^{-12}\,{\rm GeV}^{-1}$ & $<1.61\times10^{-12}\,{\rm GeV}^{-1}$  \\
\hline
\end{tabular}
\end{table*}

Uncertainties in the intensity and spectral shape of the extragalactic background light constitute an important systematic in any ALP search that relies on TeV attenuation. In the baseline analysis we adopt the Saldana-Lopez et al.\ EBL model. To assess the associated model dependence we repeated the full multi-period \(\chi^2\) profile-likelihood scan using the Franceschini et al.\ (2008) EBL model \citep{fra2008}, keeping all magnetic-field parameters fixed. The Franceschini 2008 model is chosen because it differs appreciably from the more recent Saldana-Lopez template \citep{sal2021} in the near-infrared to optical range that dominates pair-production opacity for TeV photons, and therefore provides a stringent test of the sensitivity of our constraints to EBL modelling.

For the two-zone hybrid scenario the conservative full-range limit changes from \(6.69\times10^{-12}\,{\rm GeV}^{-1}\) to \(6.80\times10^{-12}\,{\rm GeV}^{-1}\), while the limit for \(m_a<1\,{\rm neV}\) changes from \(6.35\times10^{-13}\,{\rm GeV}^{-1}\) to \(6.52\times10^{-13}\,{\rm GeV}^{-1}\). A point-by-point comparison of the two one-dimensional exclusion curves yields a mean ratio of 1.022 and an RMS logarithmic difference of 0.012\,dex; the largest local deviation is a factor of 1.11 at \(m_a\simeq45\,{\rm neV}\). For the single-zone hadronic scenario the corresponding shift is from \(7.37\times10^{-12}\,{\rm GeV}^{-1}\) to \(7.55\times10^{-12}\,{\rm GeV}^{-1}\) over the full mass range, and from \(1.57\times10^{-12}\,{\rm GeV}^{-1}\) to \(1.61\times10^{-12}\,{\rm GeV}^{-1}\) for \(m_a<1\,{\rm neV}\), with a mean ratio of 1.024 and an RMS difference of 0.011\,dex.

Although the detailed shape of the exclusion curve retains a mild dependence on the EBL choice, the one-dimensional upper limits in Table~\ref{tab:ebl} show that the most conservative bounds in both emission scenarios vary by only a few percent. The final ALP constraints are therefore not driven by the present uncertainty in the EBL model.

\section{\label{sec:Discussion}Discussion and Conclusion}
In this work, we derive the 95\% CL exclusion limits on the ALP parameters using $\chi^2$ profile likelihood analysis of five observational epochs of Mrk 421, based on simultaneous Fermi-LAT and LHAASO observations\cite{lha2026}. We treat the ALP mass $m_a$ and the photon-ALP coupling $g_{a\gamma}$ as global free parameters in a combined multi-epoch fit. To properly account for systematic uncertainties from both the observational data and the EBL model, we incorporate EBL uncertainties directly into the ALP propagation matrix and employ a Gaussian copula with correlation coefficient $\rho = 0.03$ to describe the intrinsic correlations among the multi-epoch datasets \cite{skl1959,nel2006,son2009,sat2011,joe2014,yua2018}. 

The intergalactic field enters as $g_{a\gamma}B$: a weaker IGMF loosens the upper limit on $g_{a\gamma}$. For Mrk 421 the effect is small. Raising $B_{\rm IGMF}$ from the baseline value of $0$ to $1\,\mathrm{nG}$ changes the limits by only about one percent (Table~\ref{tab:igmf}). We keep the IGMF module at $B_{\rm IGMF}=0$ rather than replacing that segment by a standalone EBL optical depth. At this redshift the two calculations nearly coincide, but they are not the same construction. Retaining the module keeps mixing and EBL absorption in one transfer matrix along the whole line of sight, and the same setup can be used without change for high-redshift sources, where a finite IGMF and in-matrix absorption are no longer interchangeable with a multiplicative EBL factor.

Under this framework, we obtain conservative 95\% CL upper limits of $g_{a\gamma} < 6.69 \times 10^{-12}\mathrm{GeV}^{-1}$ for the two-zone hybrid model and $g_{a\gamma} < 7.37 \times 10^{-12}\mathrm{GeV}^{-1}$ for the single-zone hadronic model. Additional tests performed with alternative correlation coefficients ($\rho$ = 0, 0.15, and 0.3) show that the resulting variations in the constraints remain within less than a few percentage points, indicating that the derived limits are largely insensitive to reasonable assumptions regarding weak inter-epoch correlations.

In addition, we adopt two distinct jet magnetic-field configurations to compute the photon survival probability: a two-zone hybrid model (e.g., see \cite{sah2021}) and a single-zone hadronic model (e.g., see \cite{der2012,bot2013}). The choice of these geometries follows the same observational facts already used in Sec.~I: the weak GeV--TeV correlation and the larger VHE variability amplitude favor either spatially distinct emission zones or a hadronic single zone. The field strengths themselves remain representative and are not fitted to the present SED \citep{lha2026}. Both configurations are therefore retained as alternative baselines rather than as competing best-fit jet models.

The choice of the two jet geometries follows the same observational facts already used in Sec.~I: the weak GeV-TeV correlation and the larger VHE variability amplitude favor either spatially distinct emission zones or a hadronic single zone. The field strengths themselves remain representative and are not fitted to the present SED. Both configurations are therefore retained as alternative baselines rather than as competing best-fit jet models.

For both the single-zone hadronic model and the two-zone hybrid model, the 95\% CL upper limits on the ALP-photon coupling \(g_{a\gamma}\) display a characteristic behaviour at low ALP masses. When \(m_a \lesssim 1\) neV the upper limit remains nearly constant and tight, reaching \(g_{a\gamma} < 6.35 \times 10^{-13}\,\mathrm{GeV}^{-1}\) (two-zone) and \(g_{a\gamma} < 1.57 \times 10^{-12}\,\mathrm{GeV}^{-1}\) (single-zone). An upward jump then occurs around \(m_a \approx 1\) neV, after which the upper limit enters a relatively flat plateau that extends over a wide mass range and rises only gradually with increasing \(m_a\), typically remaining at the level of a few \(\times 10^{-12}\,\mathrm{GeV}^{-1}\) up to several hundred neV.

The origin of the derived constraints is not unique across the scanned mass range and can arise from a combination of effects. At low ALP masses, the limits are primarily driven by the overall modification of the spectral shape (including possible spectral hardening or softening relative to pure EBL absorption). At higher masses, rapid photon-ALP oscillations become more prominent; however, these oscillatory features are significantly suppressed in our analysis by the Gaussian smoothing that accounts for the finite energy resolution of the instruments (adopted here as 15\%). As a result, the high-mass constraints are not solely attributable to unresolved spectral oscillations.

For a low-redshift source such as Mrk 421, an additional subtlety arises. Axion-like particles produced in the jet magnetic field may not have sufficient path length through the intergalactic magnetic field to convert efficiently back into photons before reaching the Galaxy. Combined with the relatively limited volume of the Galactic magnetic field, this can lead to a net reduction in the photon survival probability compared with the standard EBL-only attenuation. In other words, for certain regions of parameter space the presence of ALPs can produce an effective additional absorption rather than the more commonly discussed transparency enhancement (please see Fig. \ref{fig:survival_comparison}). This effect further contributes to the constraining power at intermediate and higher masses and illustrates that the limits arise from a complex interplay between jet conversion, intergalactic propagation, Galactic magnetic field effects, and instrumental energy resolution, rather than from a single dominant mechanism.

To systematically evaluate the robustness of the derived constraints, we performed dedicated re-analyses varying the jet magnetic field, IGMF, GMF model and EBL model one at a time. The resulting exclusion curves were compared via the point-by-point ratio \(R(m_a)\) and the RMS logarithmic deviation \(\mathrm{RMS}_{\log}\).

The jet magnetic field is the dominant systematic. For the two-zone model, lowering the baseline fields to \(B_{\rm in}=0.01\,\mathrm{G}\) and \(B_{\rm out}=0.001\,\mathrm{G}\) loosens the \(m_a<1\,\mathrm{neV}\) limit from \(6.35\times10^{-13}\) to \(1.63\times10^{-12}\,\mathrm{GeV}^{-1}\) (mean ratio 2.12, \(\mathrm{RMS}_{\log}=0.35\,\mathrm{dex}\)); raising them to \(B_{\rm in}=0.1\,\mathrm{G}\) and \(B_{\rm out}=0.03\,\mathrm{G}\) tightens that window to \(9.88\times10^{-14}\,\mathrm{GeV}^{-1}\). From Table~\ref{tab:jetB}, only a substantially weaker single-zone field (\(0.1\,\mathrm{G}\)) degrades the limit by about an order of magnitude (mean ratio 9.44, \(\mathrm{RMS}_{\log}=0.97\,\mathrm{dex}\)).

In contrast, Tables~\ref{tab:igmf} and~\ref{tab:gmf} show that the constraints are essentially insensitive to the IGMF (\(B_{\rm IGMF}=0\) versus the observational upper limit of \(1\,\mathrm{nG}\) yields \(\mathrm{RMS}_{\log}=0.008\,\mathrm{dex}\) and \(0.002\,\mathrm{dex}\) for the two-zone and single-zone models) and to the GMF model (Jansson \& Farrar 2012 versus Unger \& Farrar 2023, \(\mathrm{RMS}_{\log}=0.0008\)–\(0.0018\,\mathrm{dex}\)), as expected for a low-redshift, high-latitude source. The EBL template induces only a mild variation (\(\mathrm{RMS}_{\log}\approx0.011\)–\(0.012\,\mathrm{dex}\)); the one-dimensional limits in Table~\ref{tab:ebl} differ by a few percent. Overall the reported ALP limits are stable against the EBL, GMF and IGMF choices. The jet field remains the leading systematic, but it stays within a factor of a few of the baseline except for the weakest single-zone configuration.

The multi-epoch SEDs used in this work are taken from the simultaneous Fermi-LAT and LHAASO analysis of Mrk 421 presented in \citep{lha2026}. As discussed in that work, the LHAASO flux points were obtained via forward-folding and therefore retain a residual dependence on the assumed spectral shape. Systematic uncertainties on the LHAASO flux are energy-dependent and range from \(\sim\)8\% to 24\% \citep{lha2026}. In addition to the astrophysical uncertainties discussed above, we examined two instrumental systematics that are not fully captured by the published spectral energy distributions. The LHAASO flux points do not include the full energy-dependent systematic uncertainties of the instrument. To assess their potential impact we repeated the two-zone analysis after adding a conservative, energy-dependent systematic error floor that rises from 6\% at 0.1\,TeV to 15\% at 20\,TeV. The resulting most conservative upper limit weakens only mildly, from \(6.69\times10^{-12}\) to \(7.37\times10^{-12}\,\mathrm{GeV}^{-1}\) (mean ratio 1.04, \(\mathrm{RMS}_{\log}=0.026\)\,dex). The low-mass window is equally stable, shifting from \(6.35\times10^{-13}\) to \(6.55\times10^{-13}\,\mathrm{GeV}^{-1}\).

We further tested a relative energy-scale offset of \(\pm5\%\) between the Fermi-LAT and LHAASO data. A \(+5\%\) shift leaves the limits essentially unchanged (mean ratio 1.010, \(\mathrm{RMS}_{\log}=0.008\)\,dex), while a \(-5\%\) shift tightens them by approximately 2.4\% on average (mean ratio 0.977, \(\mathrm{RMS}_{\log}=0.017\)\,dex). In all cases the variation remains well within a factor of \(\sim1.3\) of the baseline result. These tests indicate that the derived ALP constraints are robust against the leading residual instrumental systematics associated with the LHAASO spectra and the inter-instrument energy calibration.

The Gaussian $\chi^2$ description assumes adequate counts in every energy bin. This is well satisfied by the Fermi-LAT points and by the lower-energy LHAASO bins, but the highest LHAASO bins have $n_{\rm eff}=(F/\sigma)^2\lesssim10$. We therefore repeated the baseline scan after discarding those bins. A point-by-point comparison of the one-dimensional 95\% C.L.\ limits yields a mean ratio of 0.979 and an RMS logarithmic difference of 0.024~dex relative to the bin-averaged baseline that retains every published point. The change is negligible below $10~\mathrm{neV}$ (mean ratio $1.001$) and remains at the percent level between $10$ and $100~\mathrm{neV}$. Only for $m_a\gtrsim100~\mathrm{neV}$ do the limits become modestly tighter (mean ratio $0.899$), as expected once the sparsely populated highest-energy bins are removed. The full-range limit changes from $6.69\times10^{-12}$ to $6.37\times10^{-12}~\mathrm{GeV}^{-1}$, while the limit for $m_a<1~\mathrm{neV}$ is essentially unchanged ($6.35\times10^{-13}$ versus $6.34\times10^{-13}~\mathrm{GeV}^{-1}$). We conclude that the low-count LHAASO bins do not drive the ALP constraints, and we retain the bin-averaged baseline that uses the complete SED as the published result.

It should be noted that the objective function employed in our $\chi^2$ profile likelihood analysis, $F=\sum \chi^2 + \text{Penalty} - 2\ln(C)$, does not strictly follow a standard $\chi^2$ distribution. Consequently, the assumptions underlying Wilks' theorem, namely independent and identically distributed data and a quadratic approximation of the likelihood surface, are not fully satisfied \citep{wil1938, cow2011}. The primary reasons are threefold: (\uppercase\expandafter{\romannumeral1}) the Gaussian copula with correlation coefficient $\rho$ introduces nonlinear coupling among the multi-epoch datasets, distorting the joint likelihood surface from a perfect paraboloid; (\uppercase\expandafter{\romannumeral2}) the nonzero correlation reduces the effective number of independent data points, thereby altering the tail behavior of the test statistic; and (\uppercase\expandafter{\romannumeral3}) the EBL penalty term effectively implements a frequentist prior that constrains the parameter space, further affecting the asymptotic properties of the test statistic near the boundaries.

Nevertheless, for the weak correlation adopted in this work ($\rho = 0.03$), the test statistic remains a good approximation to the $\chi^2$ distribution. In addition, as shown in Fig.~\ref{fig:twozone} and \ref{fig:onezone}, the coupling $g_{a\gamma}$ varies within one order of magnitude when the $\chi^2$ profile likelihood results fall between 2.3 and 20.0, while values below 2.3 yield even smaller $g_{a\gamma}$. Based on the above discussions, we adopt the conventional thresholds $\Delta F = 2.71$ and $5.99$ as a conservative phenomenological criterion for the 95\% CL upper limits. Furthermore, we enforce the monotonicity of the exclusion limits by taking the cumulative maximum of the $\chi^2$ profile likelihood, which allows for a robust and conservative treatment of statistical fluctuations. This method is widely used in high-energy astrophysics and complies with the standard analysis procedures of LHAASO and CTA (e.g., see \citep{fei2025}).

Although the jet-field strengths adopted in the baseline are not fitted to contemporaneous multi-wavelength data, they are representative values in the range commonly inferred for Mrk 421. Because the conversion probability depends on the product $g_{a\gamma}B$, changing the jet field within that range moves the low-mass bound by a factor of a few, as quantified in Table \ref{tab:jetB}. Coordinated optical-to-TeV campaigns together with the LHAASO spectra will be needed to pin down $B$ more tightly. The helical geometry assumed in the propagation is consistent with recent polarimetry of Mrk 421 (e.g,. see \citep{dig2023,kim2024}).  With these representative fields held fixed, the baseline 95\% CL limits are $g_{a\gamma}<6.35\times10^{-13}\,\mathrm{GeV}^{-1}$ (two-zone) and $g_{a\gamma}<1.57\times10^{-12}\,\mathrm{GeV}^{-1}$ (single-zone) for $m_a\lesssim1\,\mathrm{neV}$; over the full scanned range $0.1$–$500\,\mathrm{neV}$ the most conservative one-dimensional limits are $g_{a\gamma}<6.69\times10^{-12}\,\mathrm{GeV}^{-1}$ and $g_{a\gamma}<7.37\times10^{-12}\,\mathrm{GeV}^{-1}$, respectively. The same transfer-matrix setup, with mixing and EBL absorption evolved together, can be applied without change to higher-redshift sources.

\begin{acknowledgments}
The authors would like to thank the anonymous referee for the helpful comments, which improved the manuscript significantly. The authors gratefully acknowledge the financial supports from the National Natural Science Foundation of China (grants 12233006), the program for Innovative Research Team (in Science and Technology) in University of Yunnan Province (IRTSTYN), the program for Reserve Talents of Young, Middle-aged Academic and Technical Leaders in Yunnan Province (grants 202405AC350114), the Science Research Foundation of the Yunnan Education Department of China (grants NO.2024J0935, 2026J2505, 2026J2508), the Yunnan Fundamental Research Projects (grant NO. 202501AT070392), the program for Hunan Outstanding Youth Science Foundation (grants NO.2024JJ2040) and the Special Basic Cooperative Research Programs of Yunnan Provincial Undergraduate Universities' Association (grants NO.202501BA070001-122). The authors gratefully acknowledge the computing support provided by the JRT Science Data Center at Yuxi Normal University.
\end{acknowledgments}

\section*{Data Availability}
The data that support the findings of this article are openly available~\cite{lha2026}.

\nocite{*}

\bibliography{reference}

\end{document}